\documentclass[12pt,amsmath,amssymb]{article}
\usepackage[utf8]{inputenc}
\usepackage[T1]{fontenc}
\usepackage{times}
\usepackage{amsmath}
\usepackage{amssymb}
\usepackage{amsfonts}
\usepackage{amsthm}
\usepackage{mathrsfs}
\usepackage{slashed}
\usepackage{hyperref}

\newcommand{\EH}{E_H}

\newcommand{\pd}[1]{\frac{\partial}{\partial #1}}
\newcommand{\dd}[1]{\frac{d}{d #1}}
\newcommand{\ddb}[2]{\frac{d #2}{d #1}}

\newcommand{\vev}[1]{\langle #1 \rangle}

\newcommand{\matel}[3]{\langle #1|#2|#3\rangle}

\newcommand{\Lcc}{\Lambda_{\rm GT}}
\newcommand{\R}[1]{\bar{#1}}

\usepackage[all]{xy}
\begin{document}

\begin{titlepage}
\begin{flushright}\begin{tabular}{l}
Edinburgh/13/15\\
CP$^3$-Origins-2013-022\\
DIAS-2013-22
\end{tabular}
\end{flushright}

\vskip1.5cm
\begin{center}
  {\Large \bf \boldmath Renormalization group, trace anomaly and Feynman-Hellmann theorem} 
  \vskip1.3cm 
  {\sc Luigi   Del Debbio\footnote{luigi.del.debbio@ed.ac.uk}  \&
    Roman Zwicky\footnote{roman.zwicky@ed.ac.uk}}
  \vskip0.5cm
  
  $^a$ {\sl Higgs centre for theoretical physics \\
  School of Physics and Astronomy, \\
  University of Edinburgh,   Edinburgh EH9 3JZ, Scotland} \\
 
  \vspace*{1.5mm}
\end{center}

\vskip0.6cm

\begin{abstract}
  We show that the logarithmic derivative of the gauge coupling on the hadronic mass and the cosmological constant term of a gauge theory are related to the gluon condensate of the hadron and the vacuum respectively.  These relations are akin to Feynman-Hellmann relations whose derivation for the case at hand are complicated by the construction of the gauge theory Hamiltonian.  We bypass this problem by using a renormalization group equation for composite operators and the trace anomaly.  The relations serve as possible definitions of the gluon condensates themselves which are plagued in direct approaches by power divergences.  In turn these results might help to determine the contribution of the QCD phase transition to the cosmological constant and test speculative ideas.
  \end{abstract}

\end{titlepage}

\setcounter{footnote}{0}
\renewcommand{\thefootnote}{\arabic{footnote}}

\tableofcontents

\section{Introduction}

The Feynman-Hellmann theorem relates the leading order variation of the
energy to a local matrix element, providing a direct link between an
observable and a theoretical quantity.  Originally derived in quantum
mechanics, its application to quantum field theory (QFT) is generally
straightforward and widely used, see e.g.~\cite{Gasser:1979hf}.
Exceptions are cases where the Hamiltonian is difficult to construct,
which may arise as a QFT is usually defined from a Lagrangian where
(most) symmetries are manifest.  An example of such a case are gauge
theories where the elimination of two degrees of freedom from the four
vector $A_\mu$ are at the root of the problem.  In this work we bypass
the construction of the gauge field part of the Hamiltonian\footnote{In subsequent work 
the relations, obtained in this paper, have been derived using the Hamiltonian formalism \cite{Gham}.} by using
renormalization group equations (RGE) for composite operators as well
as the trace anomaly. We obtain a relation that relates the
logarithmic derivative of the hadron mass with respect to the coupling
constant, and the gluon condensate of the hadron state. Likewise we
find a similar relation relating the derivative of the cosmological
constant and the vacuum gluon condensate.

The corresponding relation for the quark mass were used in
Ref.~\cite{DelDebbio:2010jy} to derive the leading scaling behaviour of the
hadronic masses for a non-trivial infrared (IR) fixed point, deformed
by the fermion mass parameter.  The relation derived in this paper
are used to compute the scaling corrections to the hadronic mass
spectrum~\cite{size}.

\section{Preliminary results}
We shall first rederive some results before assembling them to obtain
the main relations of this paper.

\subsection{RGE for matrix elements of local composite operators} 

In this section we outline the  derivation of the standard RGE for local operator matrix elements on physical states which can be found in reference textbooks; e.g. 
\cite{ZinnJustin:2002ru,Collins:1984xc}.
We begin by defining the relation between the bare operator $O_i$ and
the renormalized operator $\R{O}_i$ 
\begin{equation}
O_i(g,m,\Lambda) =  (\R{Z}_O)_{ij}^{-1}(\mu/\Lambda) \R{O}_j(\R{g},\R{m},\mu) \;, 
\end{equation}
where summation over indices is implied, $(g,m)$ are the bare gauge coupling and mass,  
$\Lambda$ is the UV cut-off of the theory, 
$(\R{g},\R{m})$ are the set of renormalized couplings and $\mu$ is the 
renormalization scale. As stated above the operators are understood to be evaluated between two physical states in order to avoid the issue of contact terms which arises upon insertion of additional operators.
From  the independence of the bare operator on the renormalisation scale,
\begin{eqnarray}
\dd{\ln \mu} O_i(g,m,\Lambda)  = 0  \;,
\end{eqnarray}
one obtains an RGE of the form,
\begin{eqnarray} 
\label{eq:RG1}
\left(  \pd{ \ln \mu} + \R{ \beta}   \pd{\R{g}} - \R{m}\R{ \gamma}_m  \pd{\R{m}}  + ( \R{\gamma}_O)_{ij} \right) 
\R{O}_j(\R{g},\R{m},\mu)  = 0 
\end{eqnarray}
with
\begin{eqnarray}
& & \R{ \beta}  \equiv \ddb{ \ln \mu}{ \R{g}} \;, \quad  \R{\gamma}_m\equiv - \ddb{ \ln \mu}{\ln \R{m}} \;, \nonumber \\
& &  (\R{\gamma}_O)_{ij} \equiv - (\R{Z}_O^{-1})_{ik} \ddb{ \ln \mu}{ (\R{Z}_O)_{kj}} \;.
\end{eqnarray}
Denoting by $d_{O_i} \equiv d_{\R{O}_i}$ the engineering dimension of $O_i$ one gets 
by dimensional analysis 
\begin{equation}
\label{eq:RG2}
\left(  \pd{ \ln \mu}    +  \R{m} \pd{\R{m}}  - d_{O_i} \right) \R{O}_i(\R{g},\R{m},\mu)  = 0 \;,
\end{equation}
an equation which can be combined with
 \eqref{eq:RG1} into
\begin{eqnarray} 
\label{eq:RGE3}
\left( \R{ \beta}   \pd{\R{g}} - (1+ \R{ \gamma}_m ) \R{m}\pd{\R{m}}  + 
(\R{\Delta}_O)_{ij} \right) 
\R{O}_j(\R{g},\R{m},\mu)  = 0 \;.
\end{eqnarray}
Eq.\eqref{eq:RGE3} is an RGE equation for the composite operator,
sometimes referred to as the 't Hooft-Weinberg or Callan-Symmanzik
equation.  The symbol $\R{\Delta}_O \equiv d_O + \R{\gamma}_O$ is, as
usual, the scaling dimension of the operator $\R{O}$.
Eq.~\eqref{eq:RGE3} can be solved by the method of characteristics
by introducing a parameter which
has the interpretation of a blocking variable. This is for instance
used in Ref.~\cite{size} to identify the scaling corrections to
correlators at a non-trivial IR fixed point. 

To this end we note that in this paper the $O_i$ considered are physical quantities (no anomalous scaling) 
which in addition do not mix with other operators and therefore $(\Delta_O)_{ij} = d_{O_i} \delta_{ij}$.

\subsection{Trace anomaly}

Let us first define some conventions. The Lorentz-invariant
normalisation of states  for $D$-dimensional space-time is given by
 \begin{equation}
  \label{eq:normal}
  \vev{H(E',\vec{p'})|H(E,\vec{p})} = 2 E(\vec{p})   (2 \pi)^{D-1} \delta^{(D-1)}(\vec{p}-\vec{p'}) \;, 
\end{equation}
the diagonal matrix elements are abbreviated to
\begin{equation}
  \label{eq:X}
  \vev{  X }_{\EH}   \equiv     \matel{H(E,\vec{p})}{X}{H(E,\vec{p})}_c  \;,
\end{equation}
where $c$ denotes the connected part and $H$ stands for any physical
state.  Since the energy momentum tensor is related to the four
momentum operator $\hat p_\mu = \int d^{D-1} x T_{0\mu}$ it is readily
seen that:
\begin{equation}
  \label{eq:Tmunu}
  \vev{T_{\mu\nu}}_{E_H} 
  = 2 p_{\mu}  p_{\nu}\, ,  \quad \vev{T_{\mu\nu}}_{0} = \Lcc \,  g_{\mu \nu} \;,
\end{equation}
where $\vev{O}_0 \equiv \matel{0}{O}{0}$ hereafter with $| 0 \rangle$ denoting the vacuum state, $p_\mu$ is the four momentum associated with the physical state ($E = p_0$), $g_{\mu \nu}$ is the Minkowski metric with signature $(+,-,...,-)$ and $\Lcc$ is the cosmological constant contribution of  the gauge theory under consideration.

The traces of the right hand side (RHS) are the masses of the hadrons $2 M_H^2$ and 
the masses density of empty space $D \Lcc = g^{\;\; \mu}_\mu \Lcc$, and
the left hand sides (LHS) follow from the trace anomaly. For a gauge
theory with $N_f$ Dirac quarks the trace anomaly, in terms of
renormalized fields, is given by\cite{EMTtrace}\footnote{In terms of
  bare fields the expression (in dimensional regularisation) reads:
  $T_{\mu}^{\;\;\;\mu}|_{\rm on-shell} = -(D-4) \frac{1}{4} G + Q$.
},
\begin{equation} 
  \label{eq:trace}
  T_\mu^{\phantom{x}\mu}|_{\rm on-shell} = \frac{\R{\beta}}{2 \R{g}} \R{G} +  (1
  + \R{\gamma}_m)  \R{Q}  \;,  
\end{equation}
where the subscript ``on-shell'' \eqref{eq:trace} indicates that the
equation, in this form, is to be used on the physical subspace
only. Furthermore we have introduced the following shorthand notation:
\begin{equation}
\label{eq:QG}
Q \equiv N_f m \bar q q \;, \quad G \equiv \frac{1}{g^2} G^A_{\alpha \beta} G^{A\,\alpha \beta} \;,
\end{equation}
with $G^A_{\alpha \beta} $ being the gauge field strength tensor and $A$  a colour index. 
Summation over indices is understood. 
Conventions are such that  the coupling is absorbed into the gauge field.
We note that the trace energy momentum tensor is not renormalized ( i.e. $T_{\mu \nu}  = \R{T}_{\mu\nu}$)  as it is directly related 
to the four momentum which is a physical quantity. 
Furthermore $\bar Q = Q$ is an RG invariant which then implies that 
$\R{\beta}/(2 \R{g}) \R{G} + \R{\gamma}_m \R{Q}$ is an RG invariant on the subspace 
of physical states. In particular we note $\R{G} \neq G$  
which is of importance when interpreting our final result.
Finally Eqs.~\eqref{eq:trace} and \eqref{eq:Tmunu} lead to:
\begin{eqnarray}
\label{eq:2}
2 M_H^2 &=&  \frac{\R{\beta}}{2 \R{g}} \R{G}_{E_H} +  (1
  + \R{\gamma}_m)  \R{Q}_{E_H}    \\
  \label{eq:2vac}
D \Lcc  &=&   \frac{\R{\beta}}{2 \R{g}} \vev{\R{G}}_{0}  +  (1
  + \R{\gamma}_m) \vev{ \R{Q}}_0 \;.
\end{eqnarray}
 
\subsection{Feynman-Hellmann theorem in QFT}
\label{sec:FHthm}

Let us start by recalling the main steps in the derivation of the
Feynman-Hellmann theorem in quantum mechanics, which is a simple but
powerful relation which has been obtained by a number of
authors~\cite{FHclan}.  Let us consider a quantum-mechanical system,
whose dynamics is determined by a Hamiltonian $\mathcal{H}(\lambda)$,
which depends on some parameter $\lambda$. The Feynman-Hellmannn
theorem states that the $\lambda$-derivative of the energy equals the
derivative of the Hamiltonian when evaluated on the corresponding
eigenstates:
\begin{equation}
\label{eq:FH}
  \frac{\partial}{\partial \lambda} E(\lambda)   = 
  \matel{\Psi_{E(\lambda)}}{\frac{\partial}{\partial \lambda} {\cal
      H}(\lambda)}{\Psi_{E(\lambda)}} \; .
\end{equation}
It relies on the observation that 
\begin{equation}
\label{eq:essence}
  \langle \Psi_{E(\lambda)}  | \Psi_{E(\lambda)} \rangle  = 1  \quad \Rightarrow \quad 
  \frac{\partial}{\partial \lambda} \langle \Psi_{E(\lambda)}  | \Psi_{E(\lambda)} \rangle 
= 0  \;.
\end{equation}
The adaption to QFT, in the simplest cases, necessitates solely to take into account
the relativistic state normalisation \eqref{eq:normal}. 
E.g. with  $ {\cal H}_m =  Q  = \bar Q $  \eqref{eq:QG}, where ${\cal H}_{\rm tot} = {\cal H}_m + ...$, the 
Feynman-Hellmann theorem for the mass reads:
\begin{eqnarray}
\label{eq:FH1}
\R{m}  \pd{\R{m}} E_H^2 &=& \vev{\R{Q}}_{E_H} \;,  \\
\label{eq:FH1vac}
 \R{m}  \pd{\R{m}}  (D \Lcc) &=& \vev{\R{Q}}_{0} \;.
\end{eqnarray}
In \eqref{eq:FH1} we have identified $(2\pi)^{D-1}
\delta^{(D-1)}(\vec{p}-\vec{p'})_{\vec{p}\to\vec{p'}} \leftrightarrow
\int d^{D-1}x$ in the sense of distributions. Note, the $E_H^2$ instead
of $E_H$ in \eqref{eq:FH} on the LHS originates from the additional
factor of $2 E_H$ in the normalisation \eqref{eq:normal}.  In
\eqref{eq:FH1vac} the normalisation $\langle 0 | 0 \rangle = 1$ was
assumed.  Furthermore we note that in a mass independent scheme ($
\pd{m} \R{Z}_m = 0$) $\R{m} \pd{\R{m}} = m \pd{m}$ and since $Q =
\R{Q}$, therefore the relation \eqref{eq:FH1} also holds for bare
quantities. Eq.~\eqref{eq:FH1} is widely known \cite{Gasser:1979hf} 
and used in lattice simulation to extract the corresponding contribution to the nucleon
mass for example \cite{Michael}.  Noting that $m \frac{\partial}{\partial m} E_H^2 =
m \frac{\partial}{\partial m} M_H^2$\footnote{We consider the states
  as used in \eqref{eq:X} as momentum eigenstates and not boosted
  states and therefore $\vec{p}$ has no dependence on $M_H$. More
  precisely in a lattice simulation the states originate from
  interpolating operators of the form: $\Phi(\vec{p}) = \int d^{D-1} x
  e^{i \vec{x} \cdot \vec{p}}\Phi(x_0,\vec{x})$.}, it follows that
$\vev{Q}_{E_H} = \vev{Q}_{M_H}$ with normalisation \eqref{eq:normal}
is a static quantity.

\section{Gluon condensates through RGE and Feynman-Hellmann theorem}

The adaption of the analoguous relations
(\ref{eq:FH1},\ref{eq:FH1vac}) with regard to the gauge coupling $g$
is complicated by the fact that in gauge theories the construction of
the Hamiltonian itself is rather involved, see
e.g. Ref.~\cite{Ramond:1981pw}.  As stated in the introduction, we
bypass the construction of the Hamiltonian, and its primary and
secondary constraints, by using the RGE, the trace anomaly, and the
relations for the mass.

The RGE \eqref{eq:RGE3} for the $M_H^2$ and $\Lcc$\footnote{\emph{En
    passant} we note that close to the fixed point $\R{\beta} = 0$,
  the RGE for $\Lcc$ implies that $\Lcc \sim
  m^{D/(1+\gamma_m^*)}$ ($\gamma_m^*$ denotes the fixed
  point value of $\gamma_m$). This observation was essentially 
  already made in our previous
  paper \cite{DelDebbio:2010ze} by deriving $\vev{Q}_0 \sim \vev{G}_0
  \sim m^{D/(1+\gamma_m^*)}$.}\footnote{\label{foot:YM} For pure Yang Mills (YM) theory, where 
  effectively $Q \to 0$, we note that \eqref{eq:1} leads to 
  $(\R{\beta}_{\rm YM})^{-1} =  - \pd{\R{g}} \ln M_{H}$ with $H$ being a glueball state. 
   This may serve as a definition of $\R{\beta}_{\rm YM}$. 
   The extension of this idea to a gauge theory with fermions is not immediate.}
 \begin{alignat}{2} 
 \label{eq:1}
 & \left( \R{ \beta}   \pd{\R{g}} - (1+ \R{ \gamma}_m ) \R{m}\pd{\R{m}}  + 2 \right)  
M_H^2  &=& \;0  \;,  \\
\label{eq:1vac}
& \left( \R{ \beta}   \pd{\R{g}} - (1+ \R{ \gamma}_m ) \R{m}\pd{\R{m}}  + D \right)  
\Lcc \;  &=& \; 0 \;,
\end{alignat}
where $\Delta_{M_H^2} = 2$ and $\Delta_{\Lcc} = D$ are simply the
engineering dimensions as $M_H^2$ and $\Lcc$ are observables which are
free from anomalous scaling. 
Using Eqs.~(\ref{eq:1},\ref{eq:2},\ref{eq:FH1}) and
Eqs.~(\ref{eq:1vac},\ref{eq:2vac},\ref{eq:FH1vac}) we obtain:
\begin{equation}
\R{\beta} \left(  \pd{\R{g}} M_H^2 + \frac{1}{2 \bar g} \vev{\R{G}}_{E_H} \right) 
 = \R{\beta} \left(  \pd{\R{g}} \Lcc + \frac{1}{2 \bar g} \vev{\R{G}}_{0} \right) = 0
\end{equation}
and from there we read off our main results,
\begin{eqnarray}
\label{eq:final}
\bar g \pd{\R{g}} E_H^2 & =& -\frac{1}{2} \vev{\R{G}}_{E_H} \;, \\
\label{eq:finalvac}
\bar g \pd{\R{g}} \Lcc & =& -\frac{1}{2} \vev{\R{G}}_{0} \;.
\end{eqnarray}
For the first relation we have used $ \pd{\R{g}} M_H^2 = \pd{\R{g}}
E_H^2$, where the same remark applies as for the derivative with
respect to $m$ given earlier on. In particular this implies that
$\vev{\R{G}}_{E_H} = \vev{\R{G}}_{M_H}$ is a static
quantity\footnote{We could have obtained this result earlier on by
  inserting \eqref{eq:2} into \eqref{eq:FH1} and using $\vev{Q}_{M_H}
  = \vev{Q}_{E_H}$.}.  The relation remain valid in the case where the quark masses are degenerate as one can replace $m \bar q q \to m_1 \bar q_1 q_1 + m_2 \bar q_2 q_2$ in all relation as well as for the corresponding anomalous mass dimension.    
  By taking the coupling to be dimensionless we have implicitly assumed 
 the space-time dimension to be $D=4$.  
 The only modification for a  derivation in $D$-dimensions is to replace 
 $\R{\beta}/2\R{g} \to \R{\beta}/2\R{g}  + (D-4)/4$.  Since $\R{\beta}$ disappears from the final results (\ref{eq:final},\ref{eq:finalvac}), the latter are valid for any integer  $D \geq 2$.
 It seems worthwhile to point out that the
  relations (\ref{eq:final},\ref{eq:finalvac}) have been checked explicitly \cite{Gham}; 
  Eq.~\eqref{eq:final}
  for the Schwinger model and the Seiberg-Witten theory  as well as  Eq.~\eqref{eq:finalvac} for the massive multi flavour Schwinger model.
    The relations (\ref{eq:final},\ref{eq:finalvac}) (c.f. also 
  (\ref{eq:2},\ref{eq:2vac}))
  are akin to Gell-Mann Oakes Renner 
  relations \cite{GMOR}  in that they relate an operator expectation value to 
  physical quantities. 
 The scheme dependence of the gluon condensates,
inherent in the earlier statement $\R{G} \neq G$, is made manifest
through $\R{g} \pd{\R{g}} \neq g \pd{g}$ and the fact that $E_H^2$ and
$\Lcc$, being physical quantities, do not renormalize.\footnote{We wish to add the following example. 
Let $g_W(\mu)$  be a coupling defined in a Wilsonian scheme at the scale $\mu$. 
Using $g_W(\mu)$ in  \eqref{eq:FHfunM}  then  leads to a gluon condensate on the RHS
which incorporates field fluctuations up to the scale $\mu$ (in Wilsonian scheme).}  
Thus we wish
to stress that it is vital to distinguish bare and renormalized
quantities when discussing the relations (\ref{eq:final},\ref{eq:finalvac}).  
The condensates may be computed through (\ref{eq:final},\ref{eq:finalvac})
with lattice Monte Carlo simulation in some fixed scheme.  The
conversion to other schemes, say, the $\overline{\rm MS}$-scheme can
be done through a perturbative computation at some large matching
scale.  For example defining two schemes $a$ and $b$ through $g = Z_a
\R{g}_a = Z_b \R{g}_b$ one gets:
\begin{equation}
 \R{g}_a \pd{\R{g}_a} = Z_G^{ab}   \R{g}_b \pd{\R{g}_b} \;, 
 \quad Z_G^{ab}=  \left( 1+ \R{g}_b \pd{\R{g_b}}  \ln \frac{Z_a}{Z_b}  \right) \;.
\end{equation}
The transformation between scheme $a$ and $b$ is therefore given by
$\vev{ \R{G_a} } = Z^{ab}_G \vev{ \R{G_b} }$ according to
Eqs.~(\ref{eq:final},\ref{eq:finalvac}) for both the vacuum and
particle gluon matrix element.  The derivation of the relations
(\ref{eq:final},\ref{eq:finalvac}) with bare couplings would surely be
possible, but we do not consider it a necessity. 
  
It is worthwhile to illustrate the importance of using eigenstates of
the Hamiltonian for the matrix elements considered in the
Feynman-Hellmann theorem by an example at hand.  One might be tempted
to obtain the relation~\eqref{eq:FH1} directly from the trace
anomaly~\eqref{eq:trace} assuming a mass independent scheme (which
entails that $\R{\beta}$ and $\R{\gamma}_m$ are independent of $m$)
via
\begin{eqnarray}
  \R{m}  \pd{\R{m}} M_H^2 &=&  \R{m}  \pd{\R{m}}   
  \frac{1}{2} \vev{T_\mu^{\phantom{x}\mu}}_{M_H}  \nonumber \\
  &=&    \frac{1}{2}(1+   \R{\gamma}_m) \vev{\R{Q}}_{M_H}  
  +  \text{corrections} \;,
\end{eqnarray}
which without corrections and $\R{\gamma}_m \neq 1$ contradicts
\eqref{eq:FH1}.  The necessary corrections originate from the fact
that $T_\mu^{\phantom{x}\mu}$ does not commute with the Hamiltonian
in general and therefore is not an eigenoperator of the physical
states \eqref{eq:normal}. Thus differentiation of the states with
respect to $\R{m} \pd{\R{m}}$ is required for consistency and
exemplifies the importance of the energy eigenstates in the
Feynman-Hellmann theorem.

\section{Conclusions and discussion}

In this work we have derived relations between the logarithmic
derivative of the mass of a state (and the vacuum energy) with respect
to the gauge coupling in terms of the corresponding gluon
condensates as given in Eqs.~(\ref{eq:final},\ref{eq:finalvac}).
For the readers convenience we restate the relations within slightly more standard notation:
\begin{alignat}{2}
\label{eq:FHfunM}
& \R{g} \pd{\R{g}} M_H^2   &=&  -\frac{1}{2}\matel{H(E_H)}{\frac{1}{\R{g}^2}  \R{G}^2}{H(E_H)}_c    \;, 
  \\[0.1cm]
\label{eq:FHfunL}
&    \R{g} \pd{\R{g}} \Lcc &=& -\frac{1}{2} \matel{0}{\frac{1}{\R{g}^2}  \R{G}^2}{0} \;,
\end{alignat}
where $\pd{\R{g}} M_H^2 = \pd{\R{g}}E_H^2 $ as argued earlier on and  barred quantities correspond to renormalised quantities. 
In particular $\Lcc$ and $M_H^2$ originate from the trace of the  energy momentum tensor 
which is known to be finite after renormalisation of the basic parameters of the theory \cite{EMTtrace}.  Hence the relation above relate finite quantities with each other.
We shall comment on the interest 
of these equations for various aspects in the paragraphs below. 

First the $\ln \R{g}$-derivative of $M_H^2$ and $\Lcc$ may be taken as
a definition of the gluon condensates. This means that the LHS,
computable in lattice Monte Carlo simulations, serves as a definition
of the condensates on the RHS.  
An important point is that by computing the condensates indirectly via derivatives from physical quantities, problems with power divergences, which plague direct approaches, are absent.  In this respect our approach constitutes a paradigm shift.
  For $\vev{\R{G}}_{E_H}$ this should be straightforward since $M_H^2$ is easily computable whereas  
for the gluon vacuum condensate $\vev{\R{G}}_{0}$  this comes with the caveat 
that $\Lcc$ is not easy to compute by itself.  We note that for the former the disconnected part is automatically absent since it does not contribute to the mass $M_H^2$.
 The scheme dependence of the
condensate is determined by the scheme dependence of the LHS and has
been discussed in the text.  The transition from one scheme to another
can be achieved by a perturbative computation provided the matching
scale is high enough for perturbation theory to be valid.

We shall add a few remarks on  the gluon condensates.  
In QCD the matter condensates $\R{\beta}/(2\R{g})  \vev{\R{ G}}_{E_H}$ 
are known indirectly through the mass (Eq.~\eqref{eq:2}) for light mesons, other than the pseudo Goldstone bosons $\pi,K,\eta ..$,  as for the latter $\R{Q}$ is negligible since it is ${\cal O}(m_{\rm light})$.  
For the nucleon this was first discussed in \cite{Shifman:1978zn}. 
For the $B$-meson   
$\R{\beta}/(2\R{g}) \vev{\R{ G}}_{E_H}$
is related to a non-perturbative definition of the 
heavy quark scale $\overline{\Lambda}_{\rm HQ}$ \cite{Bigi:1994ga}.
The determination of the
gluon vacuum condensate is of importance for QCD sum
rules~\cite{Shifman:1978bx} as well as for the cosmological constant
problem to be discussed further below. The value of the gluon condensate cannot be regarded as settled. This is, in part,  due to the fact that 
there is no direct first principle determination of the gluon condensate.    

Let us comment on aspects of the cosmological constant, which
is a topic of more speculative nature.  Without gravity only energy differences 
matter. Thus the cosmological constant  is only determined up to a constant 
in flat space.  Yet the difference of the cosmological constant due to
the QCD phase transition itself is generally seen to be a tractable quantity,
given by Eq.~\eqref{eq:2vac} provided the condensates are well
defined.  The quark condensate is known through the Gell-Mann Oakes
Renner relation \cite{GMOR}: $\vev{\bar Q}_0 = - f_\pi^2 m_\pi^2 + {\cal O}(m) $,
with $m_\pi$ and $f_\pi$ being the pion mass and decay constants. It
would seem that any undetermined constant of the gluon condensate
should drop out in Eq.~\eqref{eq:finalvac}. Therefore $\vev{\R{G}}_0$
determined from this equation could be reinserted into
Eq.~\eqref{eq:2vac}, where scheme dependence cancels provided the
appropriate $\R{\beta}$ and $\R{\gamma}_m$ are used\footnote{A more complete discussion would include the mass degeneracy of the flavours. The effect of heavy flavours 
($m_{\rm quark} > \Lambda_{\rm QCD}$) 
can in principle be absorbed into the beta function. The precise discussion of which goes beyond the scope of this letter.}. Scheme
independence in turn might be used as a consistency check of the 
ideas brought forward in this paragraph.

Let us add that if 
 the gluon condensate can be determined, then it could be checked to
what degree the lowest $J^{\rm PC} = 0^{++}$-state in a confining
gauge theory saturates the partial dilation conserved current
hypothesis, see e.g. Ref.~\cite{Coleman}.  This could serve as a
quantitative measure to identify what is commonly referred to as a
dilaton in the literature. The possibility that the Higgs boson
candidate discovered at the LHC might be a dilaton of a gauge theory
with slow running coupling (walking technicolor) is a possibility that
is still considered within the particle physics
community e.g. \cite{dilaton-Higgs}.

It might be interesting to make use of the relation \eqref{eq:FHfunM} in 
approaches where hadron masses can be computed. 
We are thinking not only of lattice QCD but also of AdS/QCD or 
Dyson-Schwinger approaches. 
The gluon condensate could be reinserted, along with the quark condensate, 
into the trace anomaly \eqref{eq:2} and this allows for the extraction 
of information on the beta function 
and the anomalous dimension of the mass. In the case where there are either no fermions 
or fermions with zero mass, the relation in footnote \ref{foot:YM} serves as a 
definition of the beta function of the theory. 
Moroever, since the relation applies to any state one can check for the robustness of the results by applying it to many states.


{\bf Acknowledgements:} RZ acknowledges the support
of advanced STFC fellowship. LDD and RZ are supported by an STFC
Consolidated Grant.

\end{document}